\documentstyle[times,pramana,epsf,floats,psfig,graphicx,epsfig,amssymb,amsmath,latexsym,color]{ias}
\begin{document}
%\mark{{Gallant king...}{X Y zzz and A B zzzz}}
\newcommand{\M}{\mbox{m}}
\newcommand{\n}{\mbox{$n_f$}}
\newcommand{\EP}{\mbox{$e^+$}}
\newcommand{\EM}{\mbox{$e^-$}}
\newcommand{\EPEM}{\mbox{$e^+e^{-}$}}
\newcommand{\EMEM}{\mbox{$e^-e^-$}}
\newcommand{\GG}{\mbox{$\gamma\gamma$}}
\newcommand{\GE}{\mbox{$\gamma e$}}
\newcommand{\GP}{\mbox{$\gamma e^+$}}
\newcommand{\TEV}{\mbox{TeV}}
\newcommand{\GEV}{\mbox{GeV}}
\newcommand{\LGG}{\mbox{$L_{\gamma\gamma}$}}
\newcommand{\LGE}{\mbox{$L_{\gamma e}$}}
\newcommand{\LEE}{\mbox{$L_{ee}$}}
\newcommand{\LEPEM}{\mbox{$L_{e^+e^-}$}}
\newcommand{\WGG}{\mbox{$W_{\gamma\gamma}$}}
\newcommand{\WGE}{\mbox{$W_{\gamma e}$}}
\newcommand{\EV}{\mbox{eV}}
\newcommand{\CM}{\mbox{cm}}
\newcommand{\MM}{\mbox{mm}}
\newcommand{\NM}{\mbox{nm}}
\newcommand{\MKM}{\mbox{$\mu$m}}
\newcommand{\SEC}{\mbox{s}}
\newcommand{\CMS}{\mbox{cm$^{-2}$s$^{-1}$}}
\newcommand{\MRAD}{\mbox{mrad}}
\newcommand{\IND}{\hspace*{\parindent}}
\newcommand{\E}{\mbox{$\epsilon$}}
\newcommand{\EN}{\mbox{$\epsilon_n$}}
\newcommand{\EI}{\mbox{$\epsilon_i$}}
\newcommand{\ENI}{\mbox{$\epsilon_{ni}$}}
\newcommand{\ENX}{\mbox{$\epsilon_{nx}$}}
\newcommand{\ENY}{\mbox{$\epsilon_{ny}$}}
\newcommand{\EX}{\mbox{$\epsilon_x$}}
\newcommand{\EY}{\mbox{$\epsilon_y$}}
\newcommand{\BI}{\mbox{$\beta_i$}}
\newcommand{\BX}{\mbox{$\beta_x$}}
\newcommand{\BY}{\mbox{$\beta_y$}}
\newcommand{\SX}{\mbox{$\sigma_x$}}
\newcommand{\SY}{\mbox{$\sigma_y$}}
\newcommand{\SZ}{\mbox{$\sigma_z$}}
\newcommand{\SI}{\mbox{$\sigma_i$}}
\newcommand{\SIP}{\mbox{$\sigma_i^{\prime}$}}
\newcommand{\be}{\begin{equation}}
\newcommand{\ee}{\end{equation}}
\newcommand{\bc}{\begin{center}}
\newcommand{\ec}{\end{center}}
\newcommand{\bi}{\begin{itemize}}
\newcommand{\ei}{\end{itemize}}
\newcommand{\ben}{\begin{enumerate}}
\newcommand{\een}{\end{enumerate}}
\newcommand{\bm}{\boldmath}
\title{Ultimate parameters of the photon collider at the ILC~\thanks{LCWS06,
    Bangalore, India, March 2006}}

\author{V.~I.~Telnov~\thanks{e-mail: telnov@inp.nsk.su}}
\address{Budker Institute of Nuclear Physics, 630090 Novosibirsk, Russia}
%\keywords{king, minister, boredom}

\abstract{At linear colliders, the \EPEM\ luminosity is limited by
  beam-collision effects, which determine the required emittances of
  beams in damping rings (DRs).  While in \GG\ collisions at the photon
  collider, these effects are absent, and so smaller emittances are
  desirable. In present damping rings designs, nominal DR parameters
  correspond to those required for \EPEM\ collisions. In this note, I
  would like to stress once again that as soon as we plan
  the photon-collider mode of ILC operation, the damping-ring emittances
  are dictated by the photon-collider requirements---namely,
  they should be as small as possible. This can be achieved
  by adding more wigglers to the DRs; the incremental cost is easily
  justified by a considerable potential improvement of the
  \GG\ luminosity. No expert analysis exists as of yet,
  but it seems realistic to obtain a factor five  increase of the
  \GG\ luminosity compared to the ``nominal'' DR design.}

\maketitle

\vspace{-0.3cm}
\section{Introduction}
\vspace{-0.2cm}

It is well known and publicized that in addition to \EPEM\
physics, linear colliders provide a unique opportunity to study
\GG\ and \GE\ interactions at high energy and
luminosity~\cite{GKST81,GKST83,TESLATDR,TELacta1,TELacta2}.  The
physics in \GG, \GE\ collisions is very
rich~\cite{Boos,TESLATDR,Brodsky,Zerwas}.
The photon collider almost doubles the ILC physics program, while
the increase of the total cost is only a few percent.

The next few years are very important for the photon collider.
Everything that is required for the photon collider must be
properly included in the basic ILC design.  It is important to
continue the development of the physics program and start 
 the development of the laser system, which is a key element
of the photon collider. However, even more urgent are the
accelerator and interaction-region aspects, which influence the
ILC design and determine the parameters of the photon collider.

At this workshop (LCWS06), I would like to emphasize two very
important problems of the photon collider that require special
attention of ILC designers: 1) attaining the ultimate luminosities
(this article), 2) the layout of the photon collider at the ILC
~\cite{TEL-lcws06-2}.

\vspace{-0.3cm}
\section{Towards high $\boldsymbol{\gamma\gamma, \gamma e}$ luminosities}
\vspace{-0.2cm}

The \GG\ luminosity at the photon collider at ILC energies is
determined by the geometric luminosity of electron
beams~\cite{Tfrei,TEL2001,TESLATDR}.  There is an approximate general
rule: the luminosity in the high-energy part of spectrum $\LGG \sim
0.1 L_{ \rm geom}$, where $ L_{ \rm geom}=N^2 \nu \gamma / 4\pi
\sqrt{\ENX \ENY\ \beta_x \beta_y}$.  Compared to the \EPEM\ case,
where the minimum transverse beam sizes are determined by
beamstrahlung and beam instability, the photon collider needs a
smaller product of horizontal and vertical emittances and a smaller
horizontal beta-function.

The ``nominal'' (for \EPEM) ILC beam parameters are: $N= 2\times
10^{10}$, $\sigma_z=0.3$ mm, $\nu= 14100$ Hz, $\ENX = 10^{-5}$ m,
$\ENY = 4 \times 10^{-8}$ m.  Obtaining $\beta_y \sim
\sigma_z=0.3$ mm is not a problem, while the minimum value of the
horizontal $\beta$--function is restricted by chromo-geometric
aberrations in the final-focus system~\cite{TESLATDR}. For the
above emittances, the limit on the effective horizontal
beta-function is about 5 mm~\cite{TEL-Snow2005,Seryi-snow}.  The
expected \GG\ luminosity $\LGG(z> 0.8z_m) \sim 3.5 \times 10^{33}$
\CMS\ $\sim 0.17\,\LEPEM$ (here the nominal $\LEPEM = 2\times
10^{34}$ \CMS)~\cite{TEL-Snow2005}. Taking into account that cross
sections in \GG\ are larger than those in \EPEM\ collisions by one
order of magnitude, the number of events will be somewhat larger
than in \EPEM\ collisions even for the ``nominal'' case.

The above \GG\ luminosity corresponds to the beam parameters optimized
for the \EPEM\ collisions, where the luminosity is determined by
collision effects. The photon collider has no such restriction and can
work with much smaller beam sizes. The horizontal beam size at the
considered parameters (in the \GG\ case) is $\sigma_x \approx 300$ nm,
while the simulation shows that the photon collider at such energies
can work even with $\sigma_x \sim 10$ nm without fundamental
limitations~\cite{Tfrei,TEL2001,TESLATDR}. So, the nominal beam
parameters are very far from the physics limits and we should do
everything possible to minimize transverse beam sizes at the photon
collider!

Note, the minimum $\beta_x$ depends on the horizontal emittance: about
5 mm for the nominal emittance and 3.7 (2.2) mm for emittances reduced
by a factor of 2 (4), respectively~\cite{Seryi-snow,TEL-Snow2005}. In the
TESLA, emittances close to the latter case were considered:
$\ENX=0.25\times 10^{-5}, \ENY=3\times 10^{-8}$ m, which give the
\GG\ luminosity a factor of 3.5 higher!

The minimum emittances are determined by various physics effects
in damping rings such as quantum fluctuations in synchrotron
radiation and intra-beam scattering (IBS). The latter is the most
difficult to overcome. Where is the limit?  One of the possible
ways to reduce emittances is decreasing the damping time by adding
wigglers~\cite{Wolski-snow}. There are no detailed considerations
by experts yet. There are many effects in damping rings, and one
should believe only careful done studies. Nevertheless, I would
like to make some rough estimates.

The equilibrium normalized emittance in the wiggler-dominated regime
due to quantum fluctuations~\cite{Wiedemann} 
\be 
\ENX \sim 3.3 \times
10^{-11} B_0^3(\mbox{T}) \lambda_{\mathrm w}^2(\mbox{cm}) \beta_x
(\mbox{m}) \; \mbox{m}, 
\ee 
where $\lambda_{\mathrm w}$ is the wiggler
period and $B_0$ is the wiggler field (sin-like field).
 The damping time
\be
T_s = \frac{3m^2c^3}{r_e^2 E B_0^2} =
  \frac{5.2 \times 10^{-3}}{E(\GEV)B_0^2(\mbox{T})}\; \mbox{sec}.
\ee

If wigglers fill 1/3 of the DR, then for $B_0=2$ T and $E=5$ GeV
one gets $T_s=7.5\times10^{-4}$ sec, which is more than 20 times
smaller than the damping time in present designs.

For $\lambda_{\mathrm w}=10$ cm and $\beta_x=5$ m, the equilibrium normalized
emittance due to synchrotron radiation is $\ENX=1.3\times 10^{-7}$
m, which is 60 times smaller than the present nominal emittance.
The vertical emittance will be much smaller as well. 

The second effect limiting the emittance is the intra-beam scattering
(IBS). The growth time for IBS at $\E_x/\E_y$ = {\it const} depends on
the emittances roughly as $1/T_{\mathrm{\,IBS}} \sim
b/\EX^2$~\cite{Bisognano,Kubo}, where $b$ is a coefficient that
depends on the DR structure and only slightly on \EX.

In presence of both synchrotron radiation and IBS, the emittance is damped
as
\be
 \frac{d \E}{\E} \approx
-\frac{d t}{T_{s}} + \frac{\E_{s} d t }{\E\ T_s} + \frac{b d
  t}{\E^2}\, , 
\ee 
where $\E \equiv \EX$, \,\, $T_s$ is the radiation damping time, $\E_s$ the
equilibrium emittance in the absence of  IBS. This equation gives
the equilibrium emittance in the presence of IBS:
\be \E_0 = \frac{\E_s}{2} +
\sqrt{\frac{\E_s^2}{4} + b T_s}.  
\ee 
In the present DR design,  IBS adds about 20\% to
$\E_s$~\cite{Kubo} (i.e., $\E_0=1.2 \E_s$), which
gives $b T_s \sim 0.25 \E_s^2$.
For the design with $\E_s \to 0$ (see above) and a shorter  damping
time, $T_s^{\,\prime}$,  the new equilibrium emittance in IBS
dominated DR would be 
\be 
\E^{\,\prime} = \sqrt{b T_s^{\,\prime}} \sim 0.5 \E_s 
\sqrt {\frac{T_s^{\,\prime}}{T_s}}\, ,  
\ee
where the latter equality is valid only for the example above.
If we decrease the damping time by factor of 5, the resulting
emittance 
\be 
\E^{\,\prime}/\E_0 \sim  0.19.  
\ee
So, it would seem that there are a lot of resources for decreasing
the damping time and thus decreasing emittances in $x,y$
directions, as well as $\beta_x$. Until $\beta_{x,y} > \sigma_z$ 
(the hour-glass effect)  and $\sigma_y > 1$ nm~\cite{TELvert}
there is a strong dependence of the luminosity on emittances ($L
\propto 1/\sqrt{\ENX\, \ENY\, \beta_x, \beta_y}$). The decrease of the damping
time will need more RF peak power, but this problem is solvable.
The tune shift due to the beam space charge may be not important
due to strong damping.

Let us assume, as an optimistic goal, a reduction (compared to the
nominal beam parameters) of \ENX\ by a factor of 6, \ENY\ by a factor
of 4, and $\beta_x$ down to 1.7 mm (which is possible for such
emittances). Then, one can have the following parameters for the
photon collider: $N=2\times 10^{10}$, $\nu= 14$ kHz, $\ENX=1.5\times
10^{-6}$ m, $\ENY= 1.\times 10^{-8}$ m, $\beta_x=1.7$ mm,
$\beta_y=0.3$ mm, the distance between interaction and conversion
regions is 1 mm, $\sigma_x=72$ nm, $\sigma_y=2.5$ nm, $L_{\rm
  geom}=2.5\times 10^{35}$, $\LGG(z>0.8z_m) \sim 2.5\times 10^{34}$
\CMS\ $\sim 1.25 L_{e^+e^-,\, \mathrm{nomin}.}$. The resulting \GG\ 
luminosity is larger than at the nominal beam parameters by a factor
of 7.  This is very attractive and needs a serious
consideration by DR experts\,!

The (interesting) event rate in \GG\ collisions will be higher than in
\EPEM\ by one order of magnitude. This opens new possibilities, such
as the study of Higgs self-coupling in \GG\ collisions just above the
$\GG\ \to hh$ threshold~\cite{Jikia}.

\begin{figure}[hbt]
\vspace*{-0.7 cm} \hspace{-0.3cm}
\epsfig{file=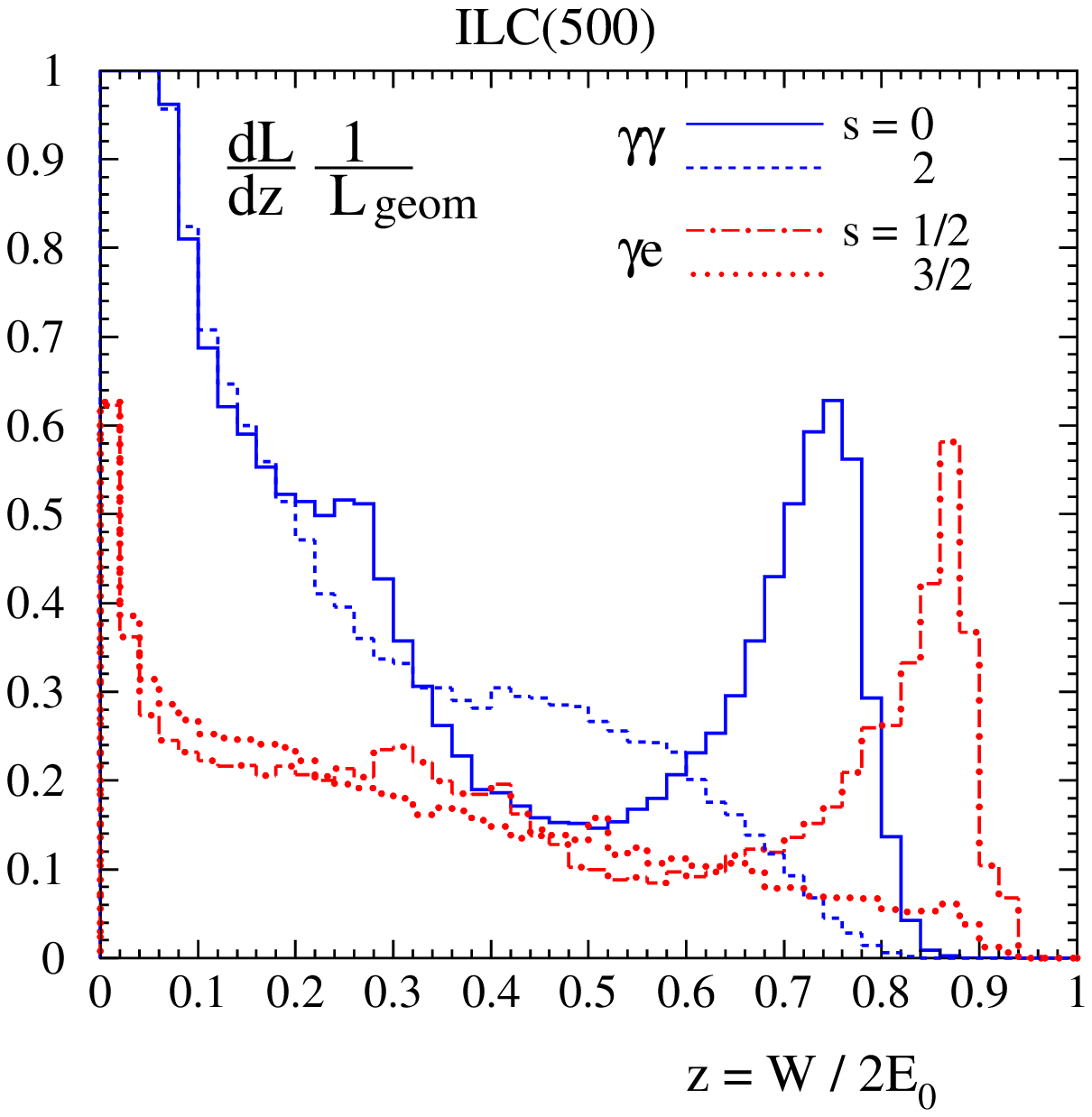,width=6.8cm,angle=0} \hspace{-1.cm}
\epsfig{file=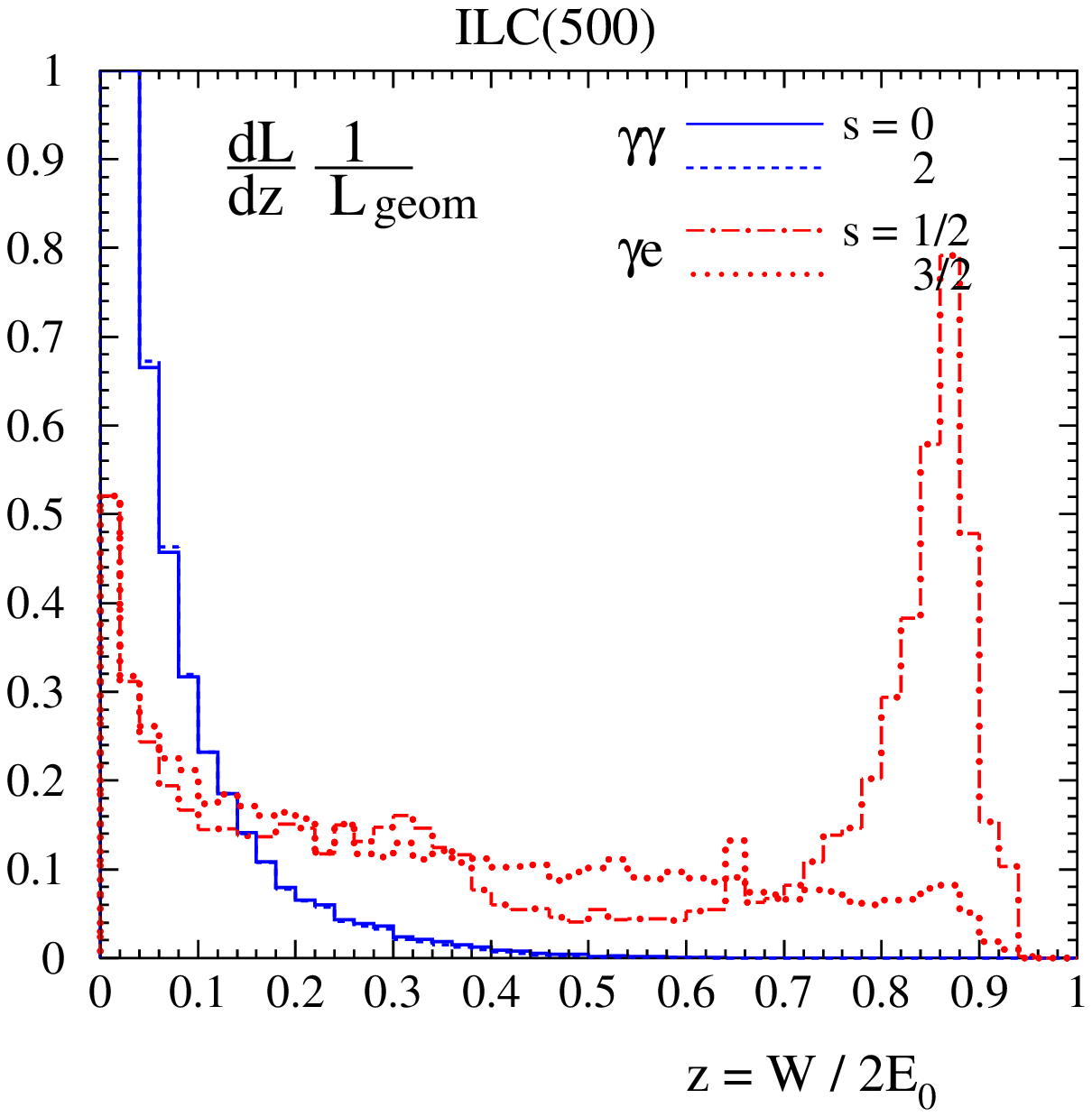,width=6.8cm,angle=0} \vspace{-0.3cm}
\caption{\GG, \GE\ luminosity spectra, left: both beams are
converted  to photons; 
\newline right: only one beam is converted to photons. See
parameters in the text. }
\label{luminosity}
%\vspace{-3mm}
\end{figure}

Figure~\ref{luminosity} shows simulated luminosity spectra for
these parameters. All important effects are taken into account. In
the figure on the right, only one of the electron beams is
converted to photons, it is more preferable for \GE\ studies due
to easier luminosity measurement~\cite{Pak} and smaller
backgrounds. The corresponding luminosities $\LGG(z>0.8z_m) \sim
2.5\times 10^{34}$ \CMS, $\LGE(z>0.8z_m)\sim 2.\times 10^{34}$ \CMS.
By increasing the distance between the conversion and interaction
regions, one can obtain a much more monochromatic luminosity
spectrum with reduced luminosity for study of QCD
processes~\cite{TEL-mont1}.

I want to stress again that parameters of the ILC damping rings
are dictated not by \EPEM, but by \GG\ collisions and a decision
on the DR design should be based on the dependence $\LGG =
f(\mbox{DR cost})$. It could be that the increase of the \GG\
luminosity by a factor of 7, as suggested above, is too
difficult, but even a 3--5 times improvement will be very useful.
This is a very important and urgent task\,!

One more remark. The photon collider does not need positrons, so
one can consider a scheme without damping rings at all.
Unfortunately, the product of emittances in polarized electron
guns is larger than in damping rings, though progress is possible.
A more radical improvement can be provided by the laser
cooling~\cite{TELlas1,TELlas2}, where intense laser beams are used
instead of wigglers. In this case, the cooling process is very
fast, there is no IBS, etc. Preliminary estimates show that the
\GG\ luminosity can be increased by a factor of 30. However, it is
too early to consider this method seriously, but it should be kept
in mind for the seconds stage of the photon collider, the ``\GG\
factory''.

\vspace{-0.3cm}
\section{Conclusion}
\vspace{-0.2cm}

In summary, in order to have a high luminosity at the photon
collider, damping rings with emittances much smaller than for
\EPEM\ are required.  No serious study has been done yet. It is
not excluded that optimized wiggler-dominated storage rings would
allow a \GG\ luminosity a factor of five higher than that in the
present design. The possibility of handling smaller horizontal
emittance should be foreseen in designs of all ILC system (bunch
compression, big bend, etc.).

\end{document}